\documentclass[twocolumn,amsmath,amssymb,superscriptaddress,nofootinbib]{revtex4}
\usepackage[english]{babel}
\usepackage[dvips]{graphicx}

\begin{document}
\title{Angular correlations in the cosmic gamma-ray background from\\
  dark matter annihilation around intermediate-mass black holes}
\author{Marco Taoso}
\affiliation{INFN, Sezione di Padova, via Marzolo 8, Padova,
35131, Italy}
\affiliation{Institut d Astrophysique de Paris, UMR 7095-CNRS,
Universit\'e Pierre et Marie Curie, 98bis boulevard Arago, 75014 Paris,
France}

\author{Shin'ichiro Ando}
\affiliation{California Institute of Technology, Mail Code 130-33,
Pasadena, California 91125, USA}

\author{Gianfranco Bertone}
\affiliation{Institut d Astrophysique de Paris, UMR 7095-CNRS,
Universit\'e Pierre et Marie Curie, 98bis boulevard Arago, 75014 Paris,
France}

\author{Stefano Profumo}
\affiliation{Santa Cruz Institute for Particle Physics and
Department of Physics, University of California, Santa Cruz, California
95064, USA}

\begin{abstract}
\noindent Dark matter (DM) annihilation could in principle
contribute to the diffuse cosmic gamma-ray background (CGB). While
with standard assumptions for cosmological and particle physics
parameters this contribution is expected to be rather small, a
number of processes could boost it, including a larger-than-expected
DM annihilation cross-section, or the occurance of DM substructures
such as DM mini-spikes around intermediate-mass black holes. We show
that angular correlations of the CGB provide a tool to disentangle
the signal induced by DM annihilation in mini-spikes from a
conventional astrophysical component. Treating blazars as a known
background, we study the prospects for detecting DM annihilations
with the Fermi Gamma-Ray Space Telescope for different choices of DM
mass and annihilation channels.
\end{abstract}

\maketitle

\section{Introduction}
\label{sec:chapter one}

The identification of nonbaryonic dark matter (DM) is still an open
problem.
Weakly interacting massive particles (WIMPs) are among the best motivated
DM candidates, due to their connections with several, independently formulated  particle physics theories beyond the Standard Model,
and also in view of their intriguing phenomenology (see
Refs.~\cite{Jungman1995,Bergstrom2000,Bertone2004,Hooper2007} for reviews).
WIMPs are actively searched for with underground detectors, with searches
for related signatures with the Large Hadron Colider, and,
indirectly, through the detection of their annihilation products.

In particular, within the latter detection technique, antimatter and gamma-ray signals
from
WIMPs annihilation are actively being searched for with the PAMELA (Payload
for Anti-Matter Exploration and Light-nuclei Astrophysics) satellite and with
the Fermi Gamma-ray Space Telescope (formerly known as GLAST), both currently taking data.
PAMELA has actually already found an interesting feature in the
positron ratio spectrum Ref.~\cite{Adriani:2008zr}, made even more interesting by the recent
ATIC data~\cite{ATIC}. This feature could be explained in terms of the
annihilation or decay of DM particles, although the gamma-ray flux
implied by these models severely constrain this interpretation
(see Ref.~\cite{Bertone:2008xr} and references therein).
Several other indirect (often conflicting) hints pointing towards
the existence of particle DM have been proposed over the last
few years, and it is therefore important to search for strategies
that allow to robustly and conclusively disentangle a DM signal
from a more mundane standard astrophysical origin.

The origin of the cosmic gamma-ray background (CGB) measured with
EGRET~\cite{Sreekumar1997} is currently uncertain and the the most
favored explanation calls for the existence of an unresolved
population of active galactic nuclei (AGNs). Recent determinations
of the gamma-ray luminosity functions (GLF) show however that
unresolved blazars alone can explain only 20-50\% of the measured
CGB~\cite{Narumoto2006}, therefore leaving room  for other gamma-ray
emitters. Besides other standard astrophysical sources, e.g.
unresolved gamma-ray emission from clusters of
galaxies~\cite{Liang2002,Rephaeli2008} or normal
galaxies~\cite{Pavlidou2002}, cosmological WIMPs annihilation could
also contribute to the CGB
\cite{Bergstrom2001,Ullio2001,Taylor2002,Elsaesser2004}.

Assuming a smooth profile for DM halos, the absence of intense
gamma-ray emission from the center of our galaxy constrain the DM
contribution from cosmological halos to be rather
low~\cite{Ando2005}, but it has been shown that the presence of
substructures can largely boost this signal without being in
conflict with galactic bounds~\cite{Horiuchi2006,Oda2005}. Here we
focus on {\it mini-spikes}, i.e. large DM overdensities that might
form around intermediate-mass black holes (IMBHs), due to the
adiabatic contraction of the DM density profile during the IMBHs'
formation and growth~\cite{Bertone2005a}. Unlike the case of a DM
spike around the central supermassive black hole (SMBH) of our
galaxy, which would inevitably be disrupted by the combination of
several astrophysical
processes~\cite{Ullio2001,Merritt:2002vj,Bertone:2005hw}, the
depletion of mini-spikes is expected to be far less efficient. It
has been shown that taking into account the contribution from
cosmological mini-spikes, DM annihilations can largely contribute to
the measured CGB~\cite{Horiuchi2006}, while spikes around SMBHs can
provide only moderate boosts~\cite{Ahn:2007ty}.

In addition, mini-spikes in the Milky May or nearby galaxies such as
M31 could be detected with neutrino telescopes~\cite{Bertone:2006nq},
boost anti-matter fluxes~\cite{Brun:2007tn}
or individually resolved by gamma-ray telescopes such as the Fermi
LAT~\cite{Bertone2005a,Fornasa2007,Fornasa:2007nr}. The simultaneous
detection of several sources with the same energy spectra, showing a
cut-off at the DM mass, would be a smoking gun for WIMPs
annihilation. On the contrary, it is difficult to extract
straightforward evidences for DM annihilation from the study of the
CGB spectrum itself. A search for these objects based on a HESS
survey of the Galactic plane region has alread allowed to set some
interesting constraints on the mini-spikes scenario
~\cite{Aharonian:2008wt}. However, additional information can be
extracted by the anisotropy data
\cite{Ando2005b,Ando2006a,Cuoco2007,SiegalGaskins2008,Hooper2007be,Lee2008}.
In particular, the CGB angular power spectrum from blazar and from
DM annihilation in halos or subhalos are quite different, due to
their different energy spectra, cosmological distribution, and
radial emissivity profiles. Therefore, the study of the CGB angular
power spectrum provides, in principle, a robust and direct tool to
discriminate between the two different scenarios. Assuming the
unresolved blazar contribution as a ``known'' background, DM
annihilation could be detected with roughly 2 years of Fermi data,
provided they contribute a fraction $\gtrsim 0.3$ of the CGB at 10
GeV~\cite{Ando2006a}.

In this paper we perform the angular anisotropy analysis for the case
of cosmological DM mini-spikes around black holes.
We compute the angular power spectrum for different DM benchmark setups,
varying the particle mass and the annihilation channel, and for
different gamma-ray energies, showing that the results are quite
sensitive to all of these variables.
We also discuss the possibility to distinguish with Fermi data the
mini-spike scenario from the case of substructure-dominated emission.

The remainder of the paper is organized as follows: Section~\ref{sec:chapter two} is
devoted to a discussion of IMBH formation and to a summary of the results of
recent numerical simulations. In Sec.~\ref{sec:chapter three}, we compute the
contribution to the CGB mean intensity from DM annihilation in cosmological mini-spikes and from
blazars, while in Sec.~\ref{sec:chapter four} we compute the angular
power spectrum for the two cases. A mixed scenario is presented in
Sec.~\ref{sec:chapter five}, where we also discuss prospects for
detecting DM annihilation with the Fermi Telescope and the effect of
changing particle DM parameters and the gamma-ray energy at which the anisotropy is studied.
Finally, conclusions are presented in
Sec.~\ref{sec:chapter six}. Throughout this paper, we adopt a flat $\Lambda$CDM model
with the cosmological parameters from WMAP 5-year data~\cite{Dunkley2008}.

\section{Intermediate Mass Black Holes}
\label{sec:chapter two}

\subsection{IMBH formation}

Black holes in the range $20 \lesssim M_{bh}/{M_{\odot}}\lesssim
10^{6}$, are commonly dubbed as IMBHs (see Ref.~\cite{Miller2003}
for a review). Despite the lack of conclusive observational evidence
for the existence of IMBH, many clues have been accumulated during
the last few years, among which the most significant is perhaps the
detection of ultra luminous X-ray sources (ULXs), interpreted in
terms of accreting IMBHs~\cite{Makishima2000,Islam2003}. Although
alternative explanations have also been proposed
\cite{King2001,Begelman2002}, they seem to be problematic or in some
cases ruled out~\cite{Miller2003,Swartz2004}, suggesting that at
least a fraction of the observed ULX sources are indeed IMBHs.
Furthermore, the presence of SMBHs at high redshifts, as indicated
by the detection of high redshift quasars at $z \sim 6$ in the Sloan
Digital Sky Survey~\cite{Fan2001,Barth2003,Willott2003}, favors a
formation scenario with rapid accretion and mergers from massive
seed BHs~\cite{Haiman2000}. Theoretically, this hierarchical scheme
would also help explain the tight observed correlations between the
SMBH mass and the host galaxy properties. IMBHs themselves could be
the remnants of first stars, also referred to as Pop III
stars~\cite{Miller2003}. These stars, the first formed in the
Universe, are generally predicted to be very massive, as a
consequence of their extremely low metallicity and the absence of
wind and pulsations. While the fate of Pop III stars with masses in
the range $100\lesssim M/M_{\odot}\lesssim250$ is perhaps to explode
due to pair instability, and to leave no compact remnant, heavier
stars finally collapse directly to IMBHs, with little mass loss.
Reference~\cite{Bertone2005a} investigated this formation scenario,
by following the evolutions of the BHs host halos and taking into
account IMBH mergers, and concluded that at $z=0$ a Milky-Way-sized
halo should host a population of roughly 1000 unmerged IMBHs, with
masses in the range $10^2-10^3$ M$_{\odot}$.

Here we consider an alternative IMBH formation scenario, proposed
in Ref.~\cite{Koushiappas2003} and further investigated in
Refs.~\cite{Bertone2005a,Koushiappas2005}. In this framework, IMBHs are
formed from gas collapsing in DM mini-halos at high redshifts. In
massive enough halos, the molecular hydrogen cooling is very
efficient and a proto-galactic disk is formed at the center.
Gravitational instabilities in the disk then trigger an effective
viscosity that drives an inflow of the gas lying in the low tail of
the angular momentum distribution. In the absence of halo mergers, the
process continues until the explosion of the first generation of
supernovae, which heats up the disk. The mass transferred to the center of
the halo undergoes gravitational collapse and a BH is then rapidly formed.
The condition that the BHs formation timescale be shorter than the
major merger timescale and that enough molecular hydrogen be present
to form a pressure support disk, sets a lower limit on the mass of
the host halo, i.e., $M_{cr}=10^{8}$ $M_{\odot}$ at the
reionization redshift $z_{re}$. The IMBH mass function is predicted to be a
log-normal, with a peak at $M_{BH}=2.1\times 10^{5}$ $M_{\odot}$
and a spread $\sigma_{BH}=0.9$.

Reference~\cite{Bertone2005a} studied the population of IMBHs that
would result from the aforementioned formation scenario in our own
Galaxy. Specifically, the authors simulated the formation of a
Milky-Way-like DM halo starting from mini-halos at high redshifts,
following the hierarchical merger history of the latter until $z=0$
in the context of a $\Lambda$CDM model for structure formation. In
that analysis, the formation of IMBHs in a given halo follows the
prescription given in Ref.~\cite{Koushiappas2003}, and pair BH
mergers occur if the pair distance is lower than 1 kpc.

IMBH formation is absent after reionization, $z<z_{re}$, since most
of the molecular hydrogen, the main baryonic coolant, is ionized. In
the simulation of Ref.~\cite{Koushiappas2005}, the authors find that
IMBHs formation is highly suppressed for $z>z_{re}$ since the
suitable hosts for BH formation become increasingly rare as redshift
increases. Therefore, according to Ref.~\cite{Koushiappas2005}, the
formation redshift distribution is peaked at $z_{re}.$

\begin{figure}[t]
\includegraphics[width=8.5cm]{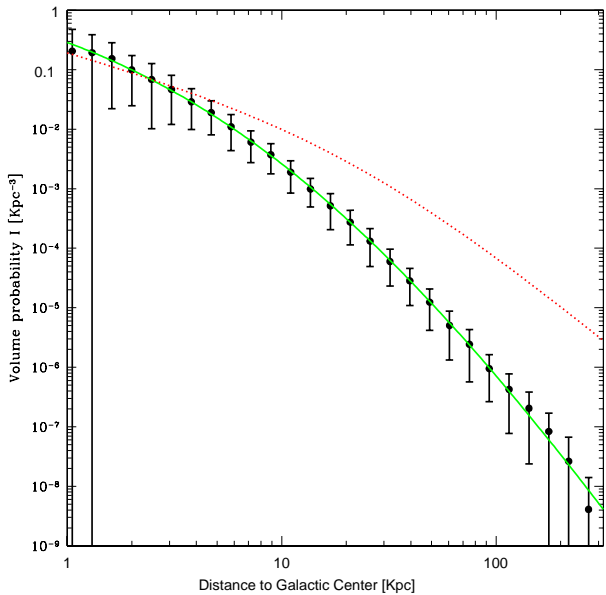}
\caption{Radial distribution of the IMBH population in the Milky
Way from the numerical results of Ref.~\cite{Bertone2005a}. The points
refer to an average among the 200 Monte Carlo realizations of the
Milky Way halo and error bars show the scatter among realizations. The
solid line is an analytical fit and the dotted line is a NFW profile.}
\label{fig:radial}
\end{figure}

The authors performed 200 statistical realizations of the IMBH
population providing for each IMBH its distance from the center of
the galaxy, its mass and the surrounding DM distributions. The
average number of unmerged IMBHs is $N_{BH}=101\pm 22$. The radial
distribution of the IMBH population is described by the volume
probability, $g(r)$, shown in Fig.~\ref{fig:radial} for an
average realization among the 200 realizations of the IMBH
population in the Milky Way. The function $g(r)$ is simply defined
as the probability to find an IMBH at a radial distance $r$ from
the Galactic center, in a spherical shell of thickness $dr$. The
volume probability function is normalized to 1 between 1 kpc and the
maximal distance from the Galactic center at which an IMBH is
found, i.e.roughly 300 kpc. The error bars in the plot reflect the
scatter among the 200 realizations.


The distribution is well fitted by the analytical function
$$g(r)=5.96 10^{-2} \left[1+\left(\frac{r}{9.1\mbox{ kpc}}\right)^{0.51} \right]^{-10.8} \mbox{ kpc}^{-3}.$$
The logarithmic slope, $\gamma= d \log g/ d \log r$, is 1.5 at 1 kpc
and 4.5 at 200 kpc, and therefore the resulting distribution is
cuspier than a Navarro-Frenk-White profile (NFW) \cite{Navarro1996},
shown in Fig.~\ref{fig:radial} for comparison.

Starting form the simulation of the IMBH population in the Milky Way
halo in Ref.~\cite{Bertone2005a}, the catalog can be adapted for other
galaxies by rescaling the total number of objects by the ratio between the
host halo masses, and the galactocentric distance by the ratio of virial
radii, $\kappa$.
The volume probability distribution function of the IMBH population in
a given galaxy, $u(r)$, is therefore obtained from that for the Milky
Way, as $u(r)=g(r/\kappa)/\kappa^3$.
This procedure has been satisfactory tested by comparing the results with a
limited number of mock catalogs, obtained as an exploratory study in
Ref.~\cite{Bertone2005a}, for different host galaxies masses.
The virial radius for a halo of a given mass $M$ at a given redshift is
defined as the radius of a spherical volume within which the mean
density is $\Delta_{c}(z)$ times the critical density at that redshift,
$M=4 \pi r_{vir}^3 \Delta_c(z) \rho_c(z)/3,$ with the virial overdensity
$\Delta_{c}(z)$ as given in Ref.\cite{Bryan1997}.

\subsection{DM distribution around IMBHs}

The process of adiabatic growth of a black hole at the center of a
DM mini-halo produces a steepening of the initial DM profile,
leading to large DM overdensities called spikes~\cite{Gondolo1999}.
Interestingly, while for the SMBHs at the center of the galaxies the
DM interaction with baryons and dynamical processes tend to weaken
the spike~\cite{Ullio2001,Bertone:2005hw}, these effects are not
effective for the dense mini-spikes around IMBHs. Starting from a DM
profile with a power law behavior in the proximity of the BH $\rho
\sim r^{-\gamma},$ the final spike profile after adiabatic growth
reads \cite{Gondolo1999}:
\begin{equation}
\rho_{sp}(r)=\rho(r_{sp})\left(\frac{r}{r_{sp}}\right)^{-\gamma_{sp}},
\label{eqn:powerlaw}
\end{equation}
where $ \rho $ is the density function of the initial profile and
the final slope $\gamma_{sp}$ is given by
$\gamma_{sp}=\frac{9-2\gamma}{4-\gamma}$, weakly depending on its
initial value $\gamma$. The radius of the gravitational influence of
the black hole $r_{h}$ sets a limit within which the spike profile
is valid, i.e. $r_{sp} \thickapprox 0.2 r_{h}$, where $r_h$ is
implicitly defined as:
$$ M(r<r_h) \equiv \int_{0}^{r_{h}} \rho(r)r^{2}dr = 2 M_{\bullet},$$
with $M_{\bullet}$ is the mass of the black hole~\cite{Bertone:2005hw}.

In the simulation of Ref.~\cite{Bertone2005a} the authors considered
an initial NFW DM density profile, with the average parameters for
the spike set as $r_{sp}=6.8\mbox{ pc}$ and $\rho_{sp}=1.2 \mbox{ }
10^{10}$ $M_{\odot}$ kpc$^{-3}.$ We employ here these reference
values throughout our analysis.


After the formation of the BH, the DM number density decreases
becuase of DM pair annihilations as: $\dot{n}_{\chi}=-(\sigma v)
n_{\chi}$ with $(\sigma v)$ the annihilation cross section times
velocity. The solution to this equation gives an upper limit to the
DM density $\rho_{lim}=m_{\chi}\times (\sigma v)^{-1} (t-t_f)^{-1}$
where $m_{\chi}$ indicates the DM mass and $t-t_f$ is the time
elapsed since BH formation. We denote $r_{lim} $ the radius where
this maximum density is reached. The density is considered to be
constant within a cut-radius defined as
$r_{cut}=\mbox{Max}[4R_{Schw},r_{lim}]$ where $ R_{Schw}$ is the BH
Schwarzschild radius. For the spikes in the simulation of Ref.
\cite{Bertone2005a}, the cut-radius at $z=0$ is, averaging over all
mini-spikes in the simulation,
 $r_{cut}=5
\times 10^{-3}$ pc, for $(\sigma v)=3$ $10^{-26}$ cm$^3$ s$^{-1}.$

\begin{figure}[t]
\includegraphics[width=8.5cm]{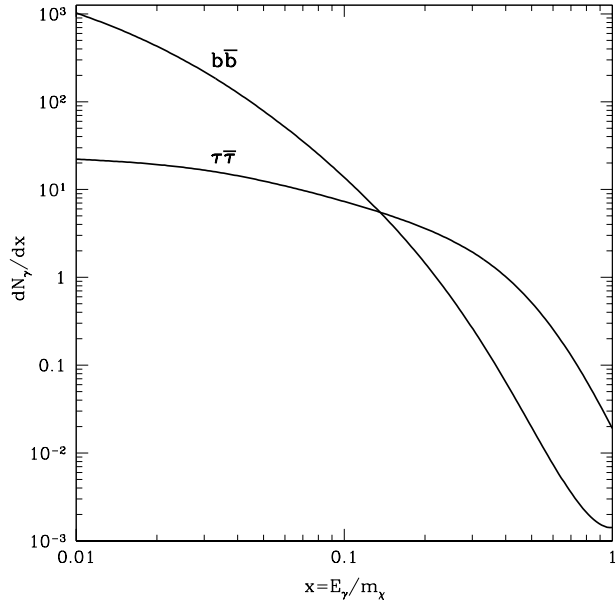}
\caption{Photon spectra for DM annihilation into $b\bar{b}$ and
$\tau^+\tau^-$. The DM particle mass is set to $m_{\chi}=100$ GeV} \label{fig:spectra}
\end{figure}
\section{Cosmic Gamma-Ray Background}
\label{sec:chapter three}

We compute in this Section the mean intensity of the CCB from unresolved mini-spikes
and from unresolved blazars.
In this and following sections, we use a similar notation to that adopted in
Ref.~\cite{Ando2006a}.

\subsection{Dark Matter Annihilations}

\subsubsection{DM annihilation contribution to cosmic gamma-ray background}

Following Ref.~\cite{Horiuchi2006}, the CGB gamma-ray flux from
cosmological DM mini-spikes, defined as the number of photons per unit
area, time, solid angle and energy, is then obtained as:
\begin{equation}
\langle I(E)_{DM}\rangle=\int dr\  W\left(E[1+z],z\right),
\label{eqn:Int}
\end{equation}
where
\begin{equation}
W(E,z)=\frac{\left(\sigma v\right)}{8\pi m_{\chi}^2}
\frac{dN_{\gamma}}{dE}\left( E[1+z]\right) e^{-\tau(E[1+z],z)}
\Delta^2(z) . \label{eqn:Intensity}
\end{equation}
The absorbtion of gamma-rays due to interaction with the diffuse
extragalactic background light is parametrized through the effective optical depth $\tau$ as in
Ref.~\cite{Bergstrom2001}.
The comoving distance $r$ and the redshift $z$ are interchangeably
used and the element $dr$ is simply $dr=c/H(z) dz$ with $H(z)$ the
Hubble function. The function $\Delta^2(z)$ in
Eq.~(\ref{eqn:Intensity}) is
$$\Delta^2(z)=n(z) \int_{r_{cut}}^{r_{sp}} \rho_{sp}^2 (r) d^3 r,$$
with $n(z)$ the comoving number density of IMBHs.

The gamma-ray annihilation spectrum $dN_\gamma/dE$ depends on the
DM particle physics model; i.e., it determines the branching ratios for annihilation in
Standard Model final states. Given these branching ratios
(which can be computed for any specified particle DM model), the quantity $dN_\gamma/dE$ can be reconstructed via Monte Carlo
simulations. This is how, for instance, $dN_\gamma/dE$ is computed in
codes like {\tt DarkSUSY}~\cite{Gondolo2004} which, in particular, makes use of {\tt
Pythia}~\cite{Sjostrand1993,Sjostrand1995} Monte Carlo simulations.

From the discussion above, it is clear that the specific DM
annihilation spectrum depends critically on the particle physics
model. In the present study we wish to consider a particle dark
matter setup as model independent as possible. As such, we consider
two representative standard model final states, and assume that the
DM particle annihilates 100\% of the time in one of those
two final states. For definiteness, we consider the final states
$b\bar b$ and  $\tau^+ \tau^-$. The choice is motivated by both
theoretical and phenomenological considerations: first, in the
context of supersymmetry, perhaps the best motivated extension to
the standard model encompassing a DM candidate, these final
states are ubiquitous; second, the resulting DM
annihilation spectra $dN_\gamma/dE$ cover the two extreme cases of a
soft photon spectrum ($b\bar b$) and of a relatively hard spectrum
($\tau^+\tau^-$). Even harder photon spectra are in principle
possible, for instance in the context of universal extra dimensions
\cite{Hooper2007}, or in other models with a large branching ratio
in charged leptons. This is not critical to us, since we only focus
on a single gamma-ray energy in our analysis; our results for the
$\tau^+\tau^-$ are conservative with respect to even harder photon
spectra, and the comparison with the soft spectrum we picked is a
solid guideline to what would change with an even harder spectrum.

In supersymmetry, in the large $\tan\beta$ regime favored by Higgs
searches at LEP, the dominant annihilation final states for the
lightest neutralino include gauge bosons (if kinematically open) and
down-type fermion-antifermion final states. The role of gauge bosons
depends on the higgsino fraction of the lightest neutralino.
Supersymmetric models with radiative electroweak symmetry breaking
and gaugino unification at the grand unification scale feature
generically a small higgsino fraction. In any case, the spectrum resulting from
gauge boson final states resembles closely the $b\bar b$ spectrum~\cite{Bertone2004,Fornengo2004,Bertone2006}.
If down-type
fermion-antifermion final states dominate, pair annihilation into
$b\bar b$ is the dominant channel, possibly competing with $\tau^+
\tau^-$ but winning over it by a factor 3 from color and by the
square of the bottom-to-tau mass ratio (see e.g. \cite{Profumo2005}). In some cases, however,
supersymmetry predicts a large branching ratio in $\tau^+ \tau^-$,
for instance when the lightest neutralino relic abundance is driven
by coannihilation with the lightest stau, which then also mediates
the dominant pair-annihilation channel. Several supersymmetric
models feature $\tau^+ \tau^-$ as the dominant annihilation
channel. In addition, other models~\cite{Hooper2007} where for
instance the quantum numbers of the DM particle weigh
favorably charged leptons over quarks, naturally feature a hard
photon spectrum, close to $\tau^+ \tau^-$.

In summary, in the present study we restrict ourselves to the two final
states $b\bar b$ and  $\tau^+ \tau^-$ as representative WIMP annihilation final states
bracketing a wide range of model-dependent predictions. The input spectra, shown in
Fig.~\ref{fig:spectra}, are the results of the numerical study of
Ref.~\cite{Cirelli2008}.

\begin{figure}[t]
\includegraphics[width=8.5cm]{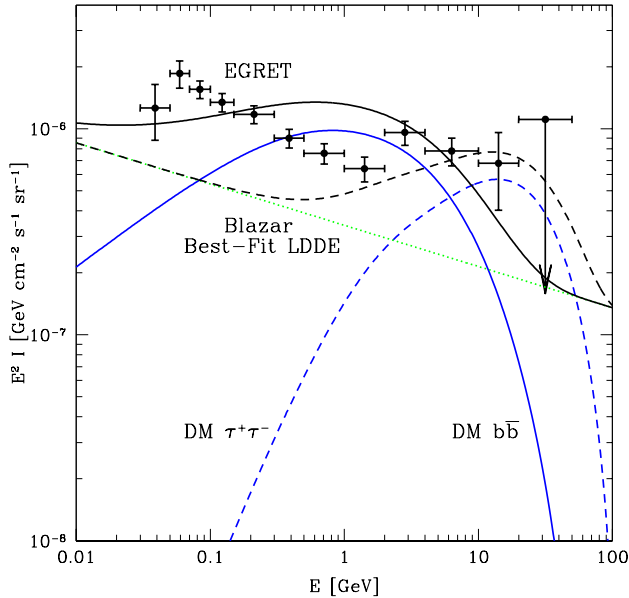}
\caption{Cosmic gamma-ray background spectrum from DM annihilation
in mini-spikes (dashed) and blazars with the best fit LDDE GLF
model(dotted). The sum of the two signals is shown as a solid line
and the data points are from EGRET data~\cite{Strong2004}.}
\label{fig:intensita}
\end{figure}

\subsubsection{Modeling the mini-spikes number density}

The IMBH number density is parametrized following the numerical
study of Ref.~\cite{Koushiappas2005}. We consider that one IMBH is
formed at redshift $z_{re}$ in every DM halo with a mass higher than
$M_{min}=10^{8} M_{\odot}.$ As mentioned in Sec.~\ref{sec:chapter
two}, IMBHs cannot be formed at more recent epochs and the formation
at higher redshifts is negligible.
In any case, the prescription we have adopted
provides a lower limit to the CGB mean flux from mini-spikes, also
in view of the fact that more than one IMBH could be formed in
larger halos as well.

The comoving number density at the formation redshift, is obtained
as:
\begin{equation}
n(z_{re})=\int_{M_{min}}^{\infty} dM \frac{dn}{dM}(M,z=z_{re}).
\end{equation}
We employ here the halo mass function $dn/dM(M,z)$ given in Ref.~\cite{Sheth1999}, with the transfer function of
Ref.~\cite{Eisenstein1997}.

After formation, IMBHs get redistributed in halos during their
hierarchical mergers. At the present epoch, the comoving number
density of unmerged IMBHs is given by:
\begin{equation}
n(0)=\int_{M_{min}}^{\infty} dM \frac{dn}{dM}(M,z=0) N_{bh}
\frac{M}{10^{12.1} h^{-1} M_{\odot}}, \label{eqn:nz0}
\end{equation}
with the average number of IMBHs in the Milky Way halo $N_{bh}$
obtained from the simulation of Ref.~\cite{Koushiappas2005}.
Here we assume a linear dependence of the
number of unmerged IMBHs on their host halo mass. As noticed in
Ref.~\cite{Horiuchi2006}, reasonable deviations from the this linear
interpolation produce small changes on the final CGB flux.

At intermediate redshift, we follow the prescription of
Ref.~\cite{Horiuchi2006}, and compute $n(z)$ assuming a redshift
power-law behavior, with the index $\beta$ obtained by fitting
$n(z)$ at $z=0$ and $z=z_{f}:$
\begin{equation}
n(z)=n(z_f)\left(\frac{1+z}{1+z_f}\right)^{\beta}.
\label{eqn:powerlaw2}
\end{equation}
Reference~\cite{Koushiappas2005} found that a Milky-Way like
galaxy would host a population of $N_{sp}=101$ IMBHs at $z=0.$ For
the same choice of the cosmological parameters and using
Eqs.~(\ref{eqn:nz}) and (\ref{eqn:powerlaw2}) we obtain $\beta=0.3,$ as
in Ref.~\cite{Horiuchi2006}. This computation can be updated by using
the more precise measurements of the cosmological parameters from
WMAP5~\cite{Dunkley2008}. Keeping $\beta=0.3,$ we obtain a sensible
decrease of the IMBH number density and for a Milky-Way like halo
at z=0, we get $N_{sp}=40.$

For the rest of the paper we will therefore assume $N_{sp}=40$ and $\beta=0.3$
to parametrize the IMBH number density.
Using Eq.~(\ref{eqn:Int}) we can now compute the mean extragalactic
gamma-ray flux from DM annihilation in cosmological mini-spikes.
The integration over $z$ is performed up to the formation redshift,
i.e., $z_{re}$. The results are shown in Fig.~\ref{fig:intensita}
adopting $m_{\chi}=100$ GeV and $(\sigma v)=3 \times 10^{-26}$ cm$^{-3}$
s$^{-1}$ and for DM annihilation into $b\bar{b}$ and $\tau^+\tau^-$.
In the same plot are shown the measurements of the CGB
extracted from EGRET data~\cite{Strong2004}. The predictions largely
depend on the annihilation spectrum, with the CGB flux peaking at
higher energies for harder spectra. For energies of the order
$\mathcal{O}(1$--$10)$ GeV, the contribution from DM annihilation is
at the same level of the CGB intensity inferred from EGRET
measurement, suggesting therefore that in this energy range DM
annihilation could substantially contribute to the total CGB flux.

In our analysis, we have included also the contribution from low
redshifts, where IMBHs are potentially detectable. Previous studies have
shown that the Fermi satellite is expected to resolve mini-spikes in our
galaxy~\cite{Bertone2005a} and maybe Andromeda~\cite{Fornasa2007} but not
further. On the other hand, the contribution of IMBHs from
$z<10^{-5}$ to the extragalactic gamma-ray background is negligible.

\subsection{Unresolved Blazars}

The gamma-ray luminosity function (GLF) of blazars is obtained from
the luminosity dependent density evolution (LDDE) model of
Ref.~\cite{Ando2006b}.

The CGB flux from unresolved blazar is computed as:
$$E\langle I_B(E) \rangle=\int_{0}^{z_{max}} dz \frac{d^2 V}{dz d\Omega}
\int_{L_{min}}^{L_{max}(z)} dL \rho_{\gamma}(L,z) \mathcal{F}_{E}(L,z).$$
The functions in the Equation above are derived in Ref.~\cite{Ando2006a}
and references therein. The minimum blazar luminosity is taken to be
$L_{min}=10^{41}$ erg s$^{-1}$ and the EGRET flux sensitivity above
100 MeV is $10^{-7}$ cm$^{-2}$ s$^{-1}.$
In Fig.~\ref{fig:intensita} we show the results for the best-fit LDDE GLF model
(details on the blazar model can be found in Ref.~\cite{Ando2006a}).

\section{Cosmic gamma-ray angular correlations}
\label{sec:chapter four}

\begin{figure}[t]
\includegraphics[width=8.5cm]{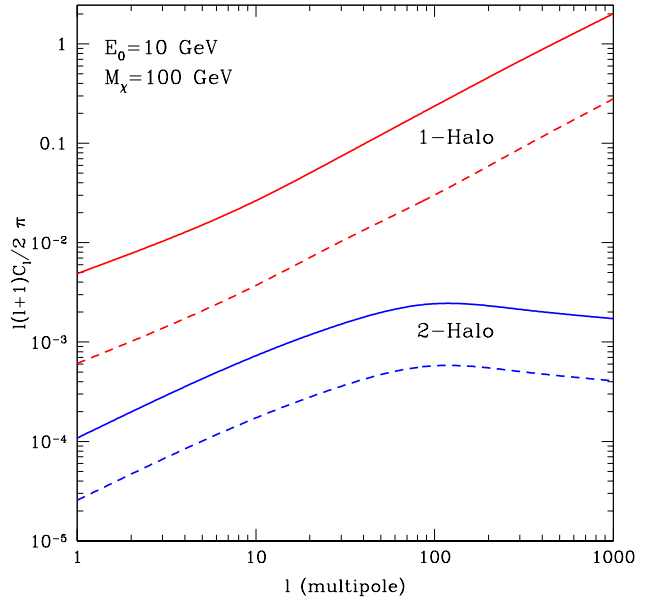}
\caption{Angular power spectrum from DM annihilation in mini-spikes
computed at $E_0=10$ GeV. We show separately the contributions from
1-halo (red) and 2-halo (blue) terms. The total angular power
spectrum is the sum of the two curves. Solid lines refer to DM
annihilation into $b\bar{b}$ and dashed ones are for
$\tau^+\tau^-$ final states. We set the DM particle mass to $m_{\chi}=100$ GeV.} \label{fig:clsdm}
\end{figure}

\subsection{Dark Matter Annihilations}

The angular power spectrum $C_l$ of the CGB from DM annihilation in
substructures has been computed in Ref.~\cite{Ando2006a}. Here we
adapt their formalism to the case of mini-spikes and we refer to the
original reference for the derivation of the equations.

The angular power spectrum from mini-spikes is obtained as:
\begin{equation}
\langle
I(E)\rangle^2 C_l=\int \frac{dr}{r^2} {W\left([1+z]E,z\right)}^2
P_{DM}\left(\frac{l}{r},z\right),
\label{eqn:Cl}
\end{equation}
where $P_{DM}(k)$ is the spatial power spectrum of mini-spikes and it can be divided
into 1-halo and 2-halo terms:
\begin{eqnarray}
P_{DM}(k)&=&P^{1h}(k)+P^{2h}(k) \label{eqn:Cl2},\\
P^{1h}(k)&=&\int_{M_{min}}^{\infty} dM \frac{dn}{dM}  \left(
\frac{\langle N|M \rangle}{n(z)} \right)^2 | u(k,M)|^2,
\label{eqn:P1}\\
P^{2h}(k)&=&\left[ \int_{M_{min}}^{\infty} dM \frac{dn}{dM}
\frac{\langle N|M \rangle}{n(z)} b(M) | u(k,M)| \right]^2 \nonumber \\
&\times&
P^{linear}(k,z). \label{eqn:P2}
\end{eqnarray}
These terms refer to correlations between two points in the same
halo (1-halo) or in two different halos (2-halo). The function $u(k,M)$ is
the Fourier transform of the IMBH volume probability, defined
in Sec.~\ref{sec:chapter two}. The linear power spectrum $P(k)$ is
obtained using the transfer function of Ref.~\cite{Eisenstein1997}
and the bias parameter is taken from Ref.~\cite{Mo1995}.

The function $\langle N |M \rangle$ gives the number of IMBHs in an
halo of given mass at a given redshift and it is related to the IMBH comoving
number density as:
\begin{equation}
n(z)=\int_{M_{min}}^{\infty} dM \frac{dn}{dM}\langle N |M \rangle.
\label{eqn:nz}
\end{equation}
As noticed in Sec.~\ref{sec:chapter two}, at $z=0$ $\langle N |M
\rangle$ is well approximated by $\langle N |M
\rangle_{lin}=N_{sp}(\frac{M}{10^{12.1} h^{-1} M_{\odot}})$ with
$N_{sp}$ corresponding to $N_{sp}=40,$ as appropriate for the Milky
Way. On the other hand, at the formation redshift we assume that one
BH is formed for every halo with mass above $M_{min}.$
Formally, there is no unique expression for $\langle N |M
\rangle$ which interpolates the two regimes above and that, at the same
time, allows one to reproduce Eq.~(\ref{eqn:powerlaw2}) from Eq.~(\ref{eqn:nz}).
We have explored different parametrization for $\langle N |M
\rangle$ encompassing its limiting behaviors at $z=0$ and $z=z_{re}$
and overestimating and underestimating $n(z)$ with respect to
Eq.~(\ref{eqn:powerlaw2}). Since the power spectrum computed in
Eq.~(\ref{eqn:Cl}) is dominated by the contribution at small $z,$ we
have found that these different choices produce differences in
$C_{l}$ always between a factor 2, that are within other
uncertainties in the calculations. This is also true also for
the cross-correlation terms that we will introduce in
Sec.~\ref{sec:chapter five}.


From Eq.~(\ref{eqn:Cl}), we note that the multipoles $C_l$ are
independent of the value of $(\sigma v)$ and of the choice of DM
density profile around each IMBH. We also find that they are weakly
dependent on the normalization $N_{bh}.$

In Fig.~\ref{fig:clsdm} we show, for the two different WIMP annihilation
channels, the contributions of 1-halo and 2-halo
terms on the angular power spectrum. We picked a gamma-ray energy at which
 we compute the anisotropy power spectrum of $E=10$ GeV, and fixed the particle DM mass to
$m_{\chi}=100$ GeV.

The 2-halo term turns out to be negligible at all angular scales.
The slope of the 1-halo term lies between those of the 1-halo terms
for annihilation in subhalos and smooth NFW halos computed in
Ref.~\cite{Ando2006a} (see their Fig.~2). This can be understood
considering that the signal in the subhalo- and in the smooth-halo-dominated
cases follow respectively the density profile and its square and
for the case of a NFW profile the two distributions are respectively
steeper and shallower than the IMBH radial distribution. The
increased normalization of the power spectrum with respect to the
case of subhalos emission is explained by the same argument: the
Fourier transform of the IMBH profile gets more power at high
$k$ with respect to that of NFW. The same tendency is found for the
two choices of the annihilation spectra. The angular power spectrum
for DM annihilation into $b \bar{b}$ is larger than that for $\tau^+\tau^-$ final states
because at the energy of $E_0=10$ GeV, the
former photon spectrum is significantly steeper than the latter.

\subsection{Blazars}

\begin{figure}[t]
\includegraphics[width=8.5cm]{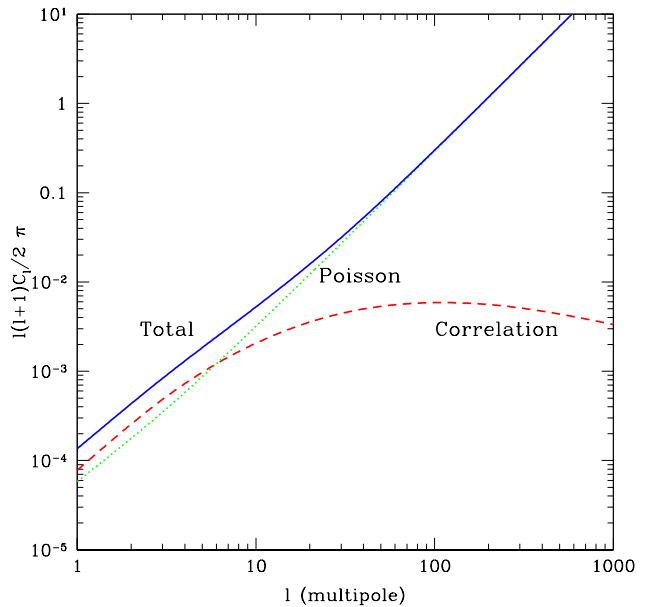}
\caption{Angular power spectrum of the CGB from unresolved blazars
expected for Fermi. We show separately the Poisson (dotted) and the
correlation (dashed) terms. The total is simply the sum, and is
shown as a solid curve. We assume here the best-fit LDDE GLF model.}
\label{fig:clblazar}
\end{figure}

\begin{figure*}[!Ht]
\includegraphics[width=8.5cm]{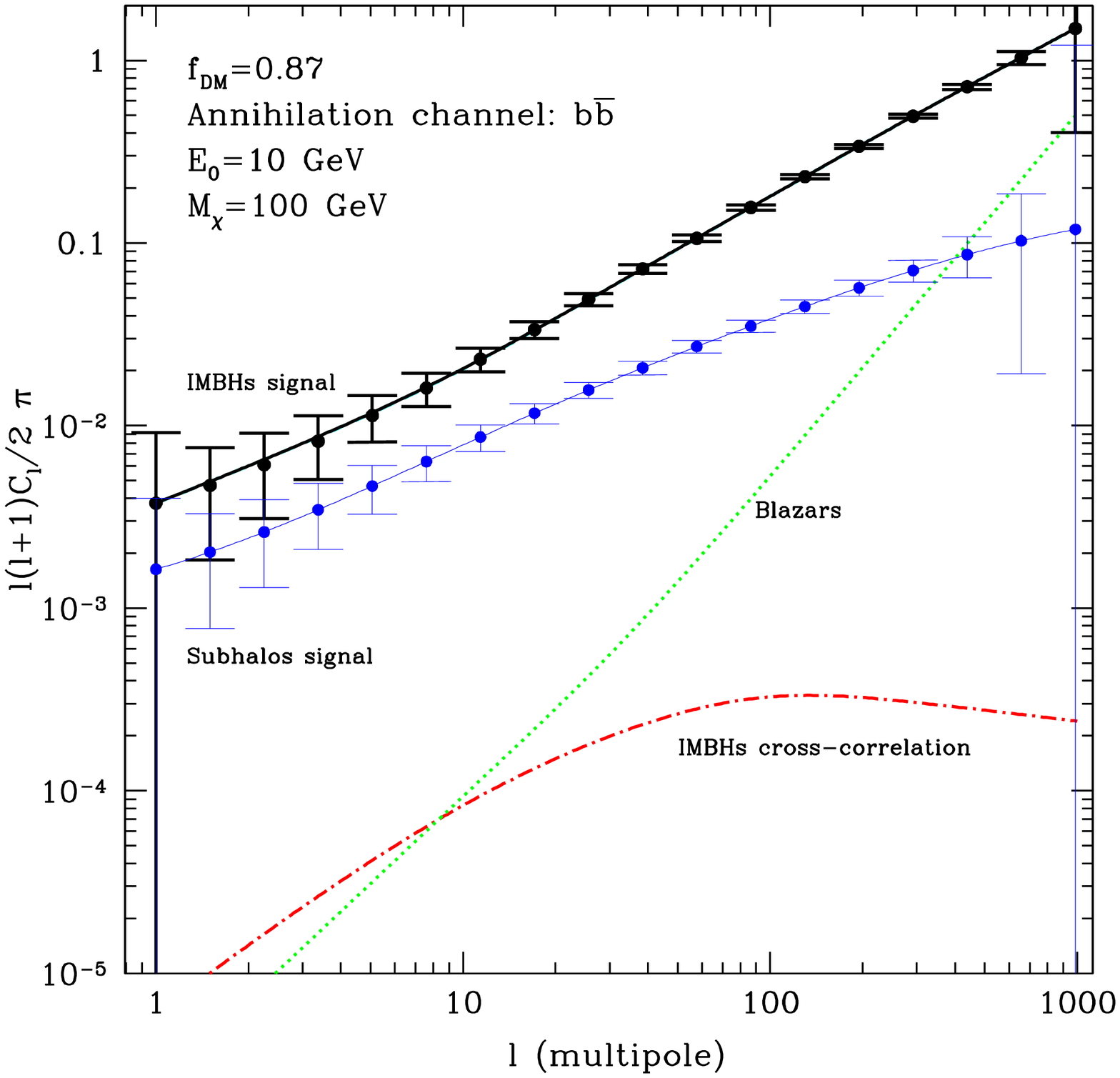}
\includegraphics[width=8.5cm]{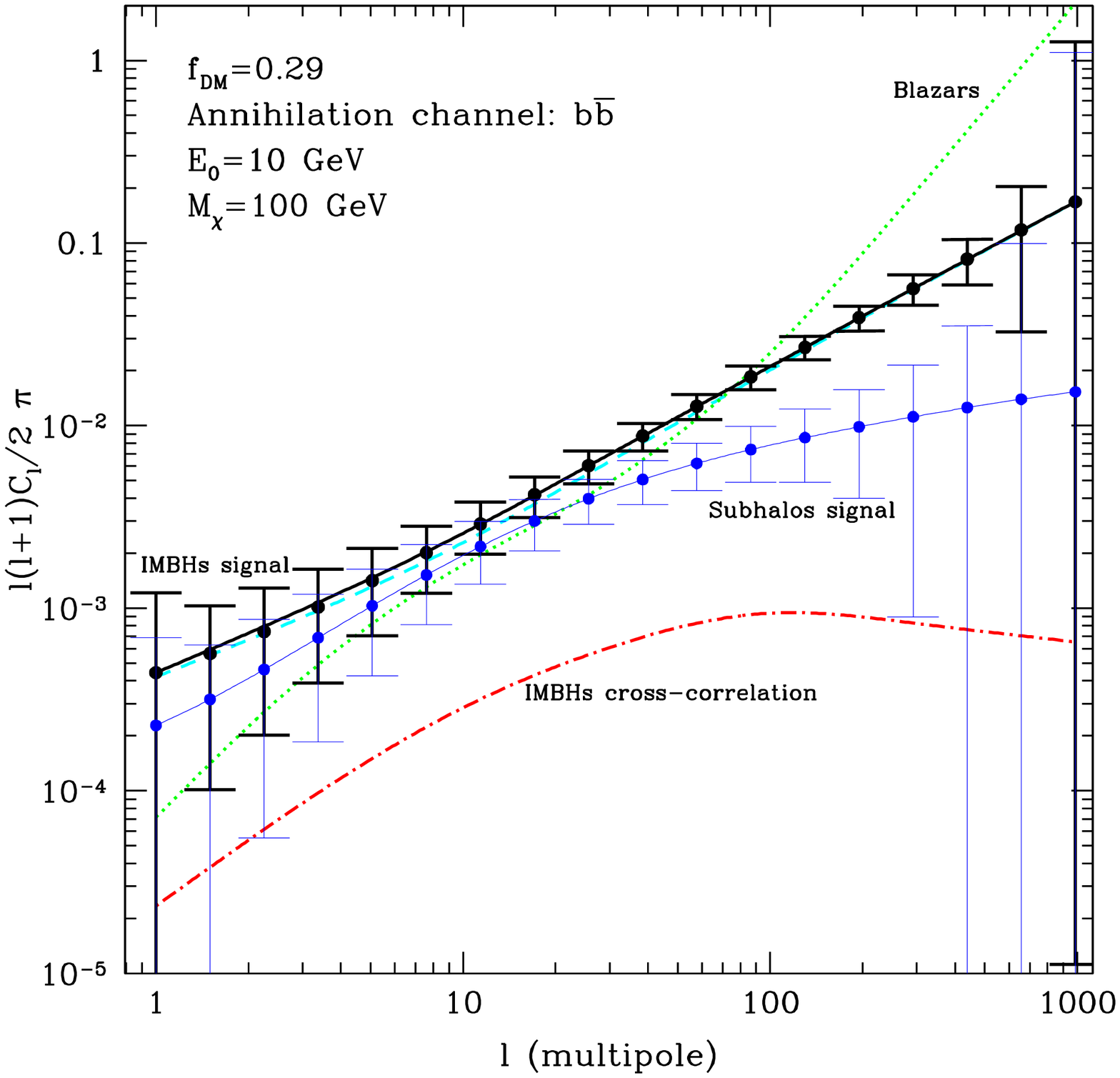}
\caption{Angular power spectrum of the CGB from DM annihilations
around IMBHs at a photon energy $E_0=10$ GeV. Dashed line shows the
contribution from DM annihilation ($f_{DM}^2C_l^{DM}$), dotted line
is for blazars ($f_{B}^2C_l^{B}$) and the dot-dashed line is the
cross-correlation term $2f_{DM}f_B C_l^{Cr}.$ The total signal
$C_l^s$ is shown as a thick black solid curve. Error bars are for
2-years of Fermi data. The thin blue solid curve show the DM signal
for DM annihilations in subhalos \cite{Ando2006a} (see text for more
details).} \label{fig:clsglast}
\end{figure*}

The angular power spectrum from unresolved blazar comes from the
contributions of a Poisson term $C_l^P$ and a correlation term
$C_l^C,$ respectively the 1-halo and 2-halo terms:
\begin{eqnarray}
C_l&=&C_l^P +C_l^C,
\label{eqn:ClBlazar}\\
C_l^P&=&\frac{1}{E^2\langle I_B(E)\rangle^2} \int dz \frac{dV}{dzd\Omega} \nonumber \\
&\times&\int^{L^{max}(z)}_{L_{min}} dL
\rho_{\gamma}(L) \mathcal{F}_E(L,z)^2,
\label{eqn:CP}\\
C_l^C &=& \frac{1}{E^2\langle I_B(E)\rangle^2} \int dz \frac{dV}{dzd\Omega} P_{lin}\left(\frac{l}{r(z)}\right) \nonumber \\
&\times& \left[  \int^{L^{max}(z)}_{L_{min}} dL \rho_{\gamma}(L)
b_{B}(L,z) \mathcal{F}_E(L,z)\right]^2 \label{eqn:CC}.
\end{eqnarray}
The blazar bias $b_{B}$ indicates how strong blazars are clustered
with compared to the linear matter power spectrum. Presently, this value
is uncertain, and different (generically inconsistent) estimates are inferred from different
techniques. Current observations give an upper bound $b_{B}\lesssim 5$
(see Ref.~\cite{Ando2006b}).

Reference~\cite{Ando2006a} estimated the correlation term
assuming either a bias model inferred from quasar observations or a
simply constant $b_{B}=1.$ The results obtained are quite similar
since the main contribution to the CGB comes from low-redshift blazars,
which have bias close to 1. In addition, for $l \gtrsim 10$ the total
angular power spectrum is dominated by the Poisson term.

We present in Fig.~\ref{fig:clblazar} our predictions for the angular
power spectrum expected to be reconstructed from Fermi data, adopting the best-fit LDDE GLF
model. We assume a Fermi point source sensitivity of $2 \times 10^{-9}$
cm$^{-2}$ s$^{-1},$  the value expected for energies
above $E=100$ MeV and two years of full sky survey mode, for sources
with a spectral index equals to 2. We note that the power spectrum
is independent of the gamma-ray energy, since we have assumed
the same power-law spectrum for all blazars and these dependence
exactly cancels when we divide by the mean intensity squared in
Eq.~(\ref{eqn:CP}) and Eq.~(\ref{eqn:CC}).

\section{Distinguishing Dark Matter Annihilation from Blazars}
\label{sec:chapter five}

We outline here the prospects for distinguishing DM
annihilation from blazar emissions in the angular power spectrum
of CGB with Fermi.

\subsection{Angular correlations of CGB in the two component case}

The CGB background receives contributions  from both DM
annihilation and from ordinary astrophysical sources, with unresolved
blazars being a representative candidate for the latter class of
emitters. For the detection of DM annihilation in the CGB,
blazars therefore constitute a background. Their contribution is
currently uncertain but we expect it will be modeled rather precisely with the Fermi catalog of detected blazars. In addition, as mentioned in
Sec.~\ref{sec:chapter four} the angular power spectrum for
astrophysical sources is energy independent and therefore it could
be calibrated at low energies where the contribution from DM
annihilation is negligible and then subtracted from the
total anisotropy data.
For this analysis we therefore treat the blazar contribution as a
known background, and we study the prospects for detecting DM
annihilation on top of it.

In this two component analysis, the total CGB intensity is the sum
of the DM and blazar contributions:
$$\langle I_{CGB}(E)\rangle=\langle I_{DM}(E)\rangle+\langle
I_{B}(E)\rangle.$$
Labeling with $f_{DM}$ the fraction of the total CGB coming from DM
annihilation, $f_{DM}=\langle I_{DM}(E)\rangle/\langle
I_{CGB}(E)\rangle,$ the total angular power spectrum is:
$$C_l^{CGB}=f_{DM}^2 C_l^{DM}+2 f_{DM}f_BC_l^{Cr}+f_{B}^2 C_l^{B},$$
where $C_l^{DM}$ and $C_l^{B}$ are respectively the angular power
spectrum from DM annihilation and blazars and $f_B$ is
simply $f_B=1-f_{DM}.$

The cross-correlation term $C_l^{Cr}$ has been studied in
Ref.~\cite{Ando2006a} and is divided into 1-halo and 2-halo terms:
\begin{eqnarray}
C_{l}^{Cr,1-halo}&=&\frac{W\left([1+z]E,z\right)}{E\langle I_B(E)\rangle \langle I_{DM}(E)\rangle} \int^{L^{max}(z)}_{L_{min}} dL \rho_{\gamma}(L)\nonumber \\
&\times& \mathcal{F}_E(L,z) \frac{\langle N|M \rangle}{n(z)} u\left(\frac{l}{r},M[L] \right),
\label{eqn:Ccross1h}\\
C_l^{Cr,2-halo} &=&\frac{W\left([1+z]E,z\right)}{E\langle I_B(E)\rangle \langle I_{DM}(E)\rangle}
\int^{L^{max}(z)}_{L_{min}} dL \rho_{\gamma}(L)\nonumber \\
&\times& \mathcal{F}_E(L,z) b_{B}(L,z) \int_{M_{min}}^{\infty} \frac{d n(M,z)}{d z)} \frac{\langle N|M \rangle}{n(z)} \nonumber \\
&\times& b(M,z) u\left(\frac{l}{r},z,M\right) P_{lin}\left(\frac{l}{r},z\right).
\label{eqn:Ccross2h}
\end{eqnarray}
A relation between the blazar luminosity and its host halo mass, $M[L]$ is given
in Ref.~\cite{Ando2006b}.

In this two component framework, the total signal $C_l^s$ and the
background noise $C_l^b$ therefore read:
\begin{eqnarray}
C_l^s&=&f_{DM}^2+2f_{DM}(1-f_{DM})C_l^{Cr},\\
C_l^b&=&(1-f_{DM})^2 C_l^{B}. \label{eqn:Cls}
\end{eqnarray}

The GLF-LDDE blazar model in Ref.~\cite{Ando2006b} basically depends
on three parameters $(\gamma_1,q,k)$  and as reminded in
Sec.~\ref{sec:chapter three}, the best-fit model only accounts for
$\sim$15\% of the CGB intensity at 10 GeV. However, varying the
parameters of the blazar model allows to explain different fractions
of the CGB. For example setting them to ($\gamma_1=1.36$, $q=3.80$,
$k=3.15 \times 10^{-6}$) we obtain a blazar fraction $f_B=0.71.$ On
the other hand, the contribution from DM annihilation in mini-spikes
is largely affected by astrophysical and particle physics
uncertainties. For example, in WIMP models the mass usually lies in
the broad range $\mathcal{O}(1$--$1000)$ GeV\footnote{For a recent
discussion of ultra-light WIMPs in supersymmetry see
Ref.~\cite{Profumo2008}.} and $( \sigma v)$ can largely differ from
the thermal value $( \sigma v )=3 \times 10^{-26}$ cm$^3$ s$^{-1}$
in the presence of efficient coannihilations or Sommerfeld
corrections, if the DM candidate is nonthermally produced or if a
modified cosmological expansion rate is postulated at the time of
WIMP freeze-out. Moreover, as discussed in Sec.~\ref{sec:chapter
three}, the number of mini-spikes in halos could differ from that
found in simulations, since IMBH formation could have been
underestimated or on the contrary IMBHs could have been more
efficiently destroyed by astrophysical processes than what is
expected. In addition, the DM density profile around each IMBH could
be modified as well by feedback.

Motivated by these arguments we compute two different
models of blazars, explaining respectively a small and an high
fraction of the CGB at 10 GeV, and we assume that the remaining CGB
intensity comes from DM annihilation in mini-spikes. As a benchmark
model we fix the DM mass to $m_{\chi}=100$ GeV and we refer to
annihilation to $b \bar{b}$ pairs. Following Ref.~\cite{Ando2006a}, we
choose an energy of observation $E_0=10$ GeV as a compromise between
maximization of signal count and minimization of the Galactic
emission. At lower energies, the galactic foreground becomes stronger,
masquerading the extragalactic component, while at higher energies,
the photon count is more suppressed. However, we perform
our analysis also for different choices of DM parameters
and energies of detection.

\subsection{Prospect for detection with the Fermi Telescope}

The Large Area Telescope (LAT) onboard the Fermi satellite is currently
taking scientific data in a survey mode. The LAT has a more than one
order of magnitude better sensitivity in the 20 MeV to 10 GeV region
than its predecessor Energetic Gamma Ray Experimental Telescope (EGRET)
onboard the Compton Gamma-ray Observatory~\cite{Cgro}. In addition, the LAT
extends the high-energy gamma-ray region up to
around 300 GeV. In the present study, we consider a mean exposure of
$1.2\times 10^{11}\ {\rm cm}^2\ {\rm s}$, corresponding, roughly, to
2 years of all-sky survey mode operation~\cite{Atwood1993,Michelson2007,Glast}.
We assume an angular
resolution for 68\% containment of the point spread function of
$\sigma_b=0.115^\circ$, appropriate for energies of around 10 GeV.
Our choices reflect those described in Ref.~\cite{Ando2006a}. The
angular resolution improves at larger energies, and degrades at
lower energies.

For the type of study hereby presented, a thorough knowledge of the
gamma-ray galactic background will be warranted. In addition,
disentangling the diffuse extra-galactic background from the
mentioned galactic emission will also be challenging. Realistically,
the 2 years of observations we consider refer not to the early
stages of the mission but, rather, to a stage when these backgrounds
are considered to be thoroughly under control.

Considering the Fermi specifications described above, the projected
1-$\sigma$ error bars of the CGB power spectrum from DM
annihilation is:
\begin{eqnarray}
\delta C_l^s=\sqrt{\frac{2}{(2l+1)\Delta l f_{sky}}} \left(
C_l^s+C_l^b+\frac{C_N}{W_l^2}\right).
\label{eqn:Clerror}
\end{eqnarray}
We take a bin width $\Delta l=0.5 l.$ The window function of a
gaussian point spread function is $W_l=\exp (-l^2 \sigma_b^2/2).$
$C_N$ is the photon spectrum of the photon noise and it is given by
$C_N=\Omega_{sky}N_{tot}/N_{CGB}^2$ with $N_{tot}$ and $N_{CGB}$
respectively the total and CGB photon numbers detected from a region
of sky $\Omega_{sky}.$

Following Ref.~\cite{Ando2006a}, we restrict the analysis to galactic
latitudes $|b|>20^{\circ}.$ At lower latitudes, the
galactic foreground dominates over the CGB flux, while the
situation is expected to be reversed in the region we consider. After
the cut of the galactic plane, the fraction of sky we consider is
$f_{sky}=0.66$. Using $N_{tot}\sim N_{CGB}$ we obtain $C_N\sim 4\pi
f_{sky}/N_{CGB}=8 \times 10^{-5} (E/10\mbox{ GeV}).$
Here we employ the total CGB flux as estimated from EGRET data.
We note, however, that since Fermi is expected to detect a large number of
blazars, the total GCB intensity will be in all likelihood reduced, possibly lowering our
error estimations.

In Fig.~\ref{fig:clsglast}, we present our predictions for two blazar
models contributing a fraction $f_B=0.13$ and $f_B=0.71$ of the total CGB flux
at $E_0=10$ GeV.
We show the signal and the background power spectra that Fermi is
expected to measure after two years of observations as well as the
projected 1-$\sigma$ signal error bars. The signal is detected if
$C_l^s >\delta C_l^s.$ We notice that this occurs even if the DM
contribution is very small. In addition, the shape of the DM power
spectrum is very different from the one corresponding to blazars. This
feature could therefore help distinguish the two scenarios.

In Refs.~\cite{Ando2005b,Ando2006a}, the authors first studied the
angular anisotropies of the CGB from DM annihilation. They focused their
attention on two scenarios, the first assuming that the DM signal is
dominated by annihilations occurring in cosmological DM halos and
the latter considering that the dominant contribution comes from the
populations of DM clumps hosted in the main DM halos. For each
possibility they took into account the possible uncertainties on the
minimum halo mass value and on the halo occupation distribution,
i.e., the number of subhalos in a parent halo of given mass. For
these frameworks, they computed the angular power spectrum from DM
annihilation that Fermi is expected to measure. They concluded that
provided DM annihilation contribute to the CGB at 10 GeV with a
fraction $f_{DM}\gtrsim0.3,$ after two years of data taking Fermi
will be able to detect the DM signal.

In Fig.~\ref{fig:clsglast}, we show their results for the most
promising case, i.e., when the DM signal is dominated by
cosmological clumps with an halo occupation distribution $\langle
N|M\rangle \propto M.$ We consider that DM annihilations in subhalos
contribute to a certain fraction $f_{DM}$ to the CGB intensity at 10
GeV and the remaining flux comes from blazars. In each plot the DM
fraction $f_{DM}$ is the same for the mini-spike and clump
scenarios. The signal for the subhalo case with the associated
1-$\sigma$ error bars is plotted as a thin blue solid line.
Comparing the DM signals in the mini-spike and subhalos scenarios we
notice that provided that DM annihilations largely contribute to the
CGB mean intensity, there are promising prospects for distinguish
the two cases. This conclusion is further strengthened if we
consider DM annihilation in cosmological smooth halos instead of
clumps, since, as stated before, the expected angular power spectrum
is smaller than when subhalos emission dominates.

\begin{figure*}[t]
\includegraphics[width=8.5cm]{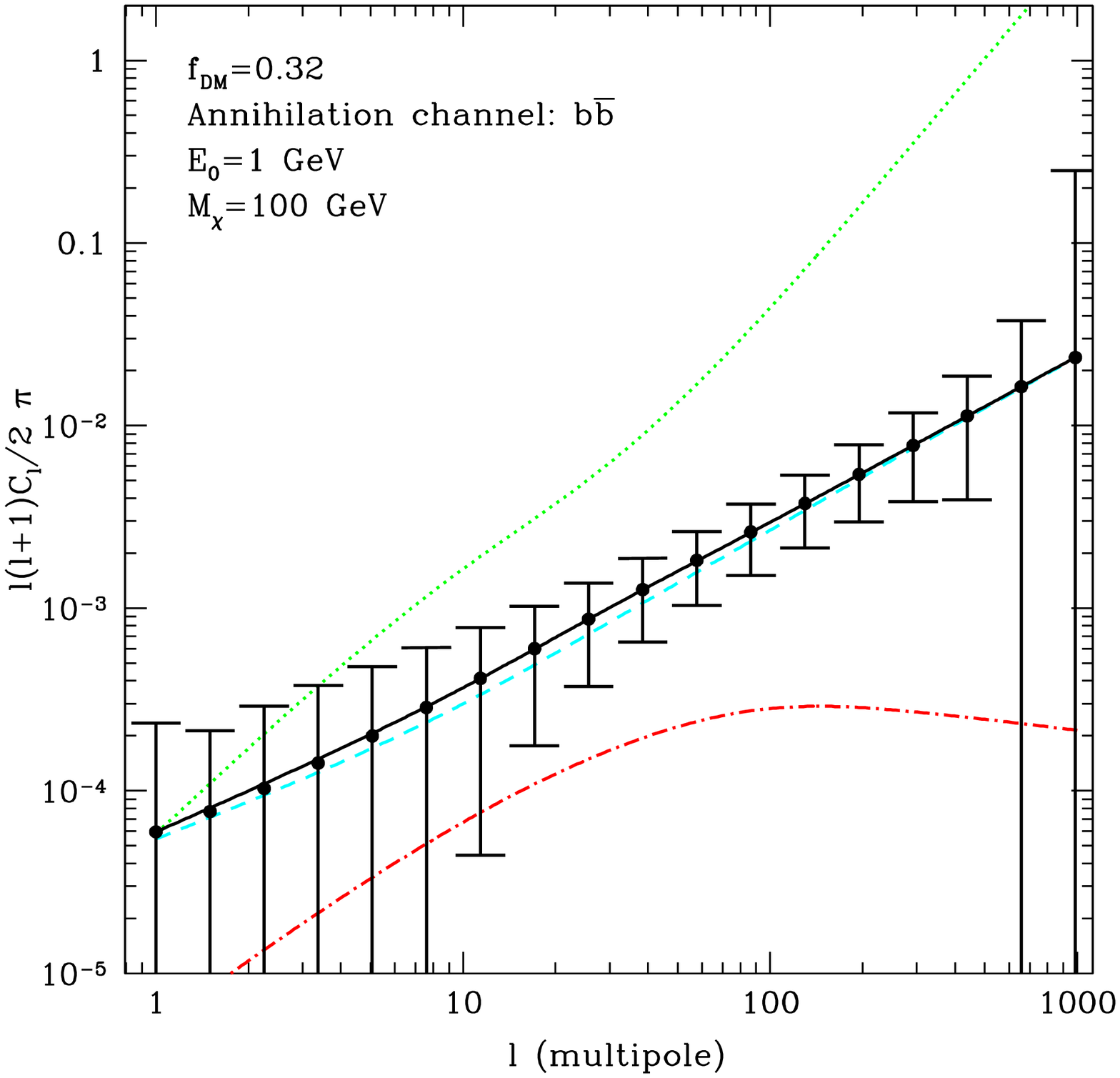}
\includegraphics[width=8.5cm]{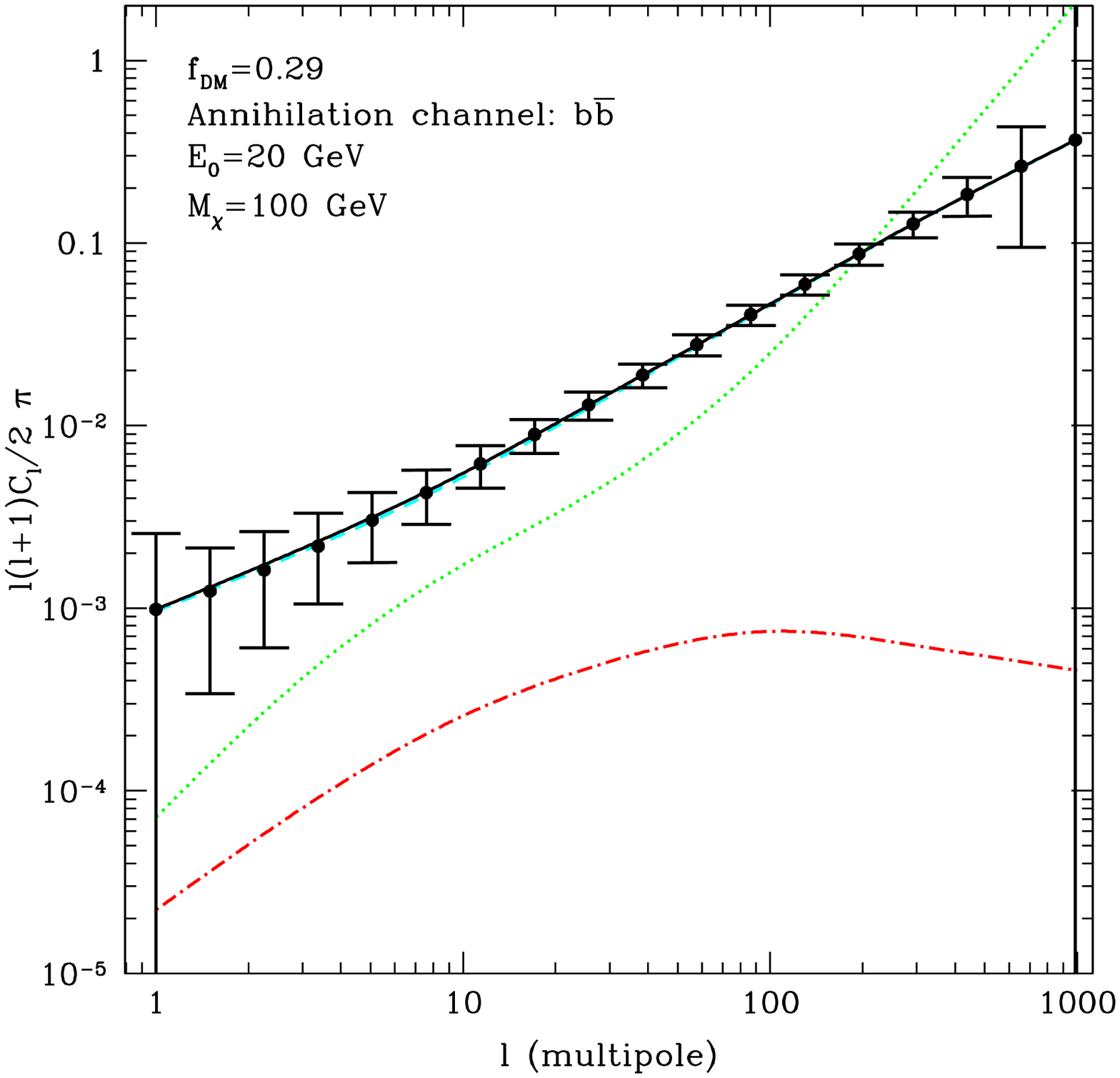}
\includegraphics[width=8.5cm]{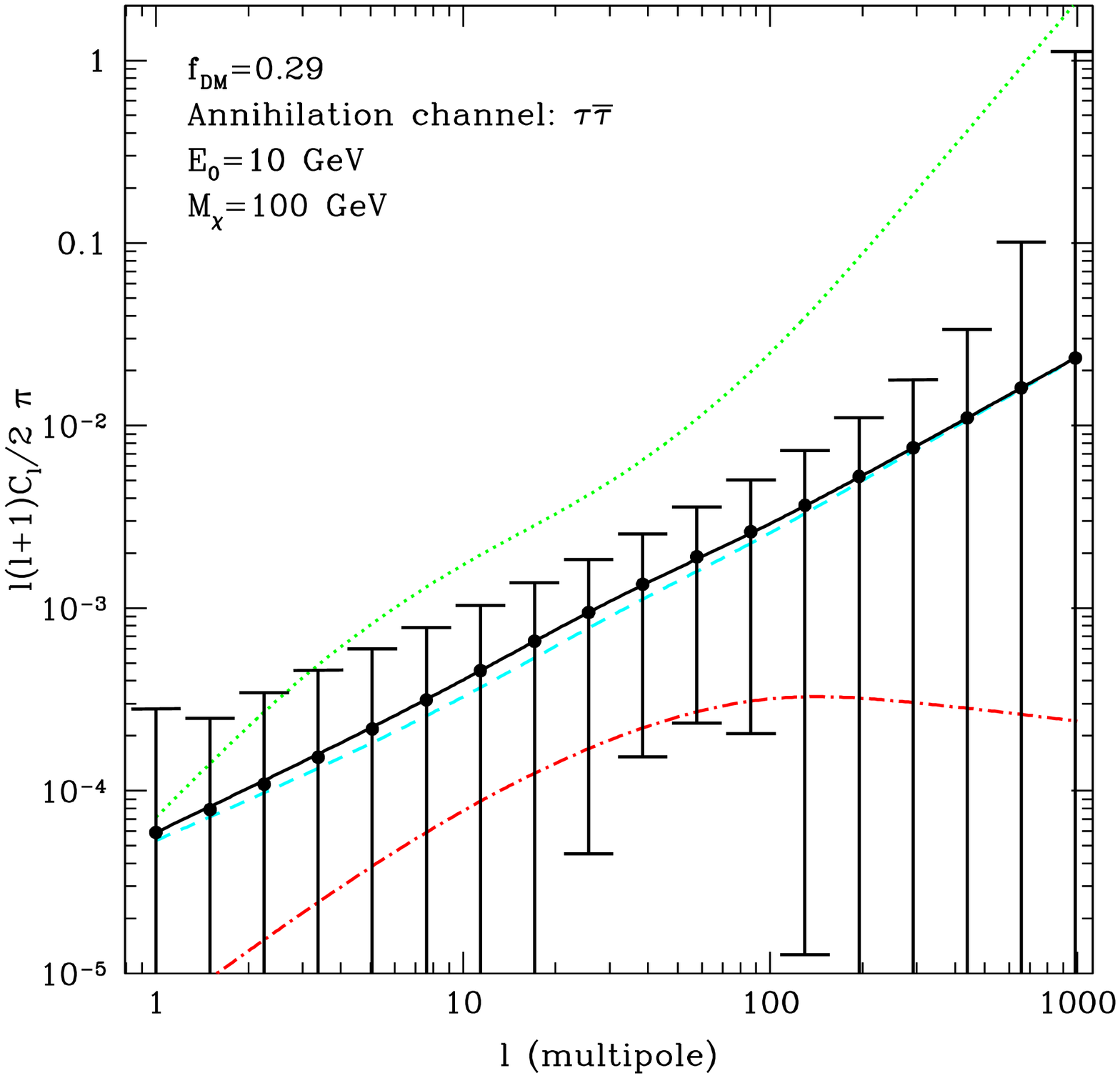}
\includegraphics[width=8.5cm]{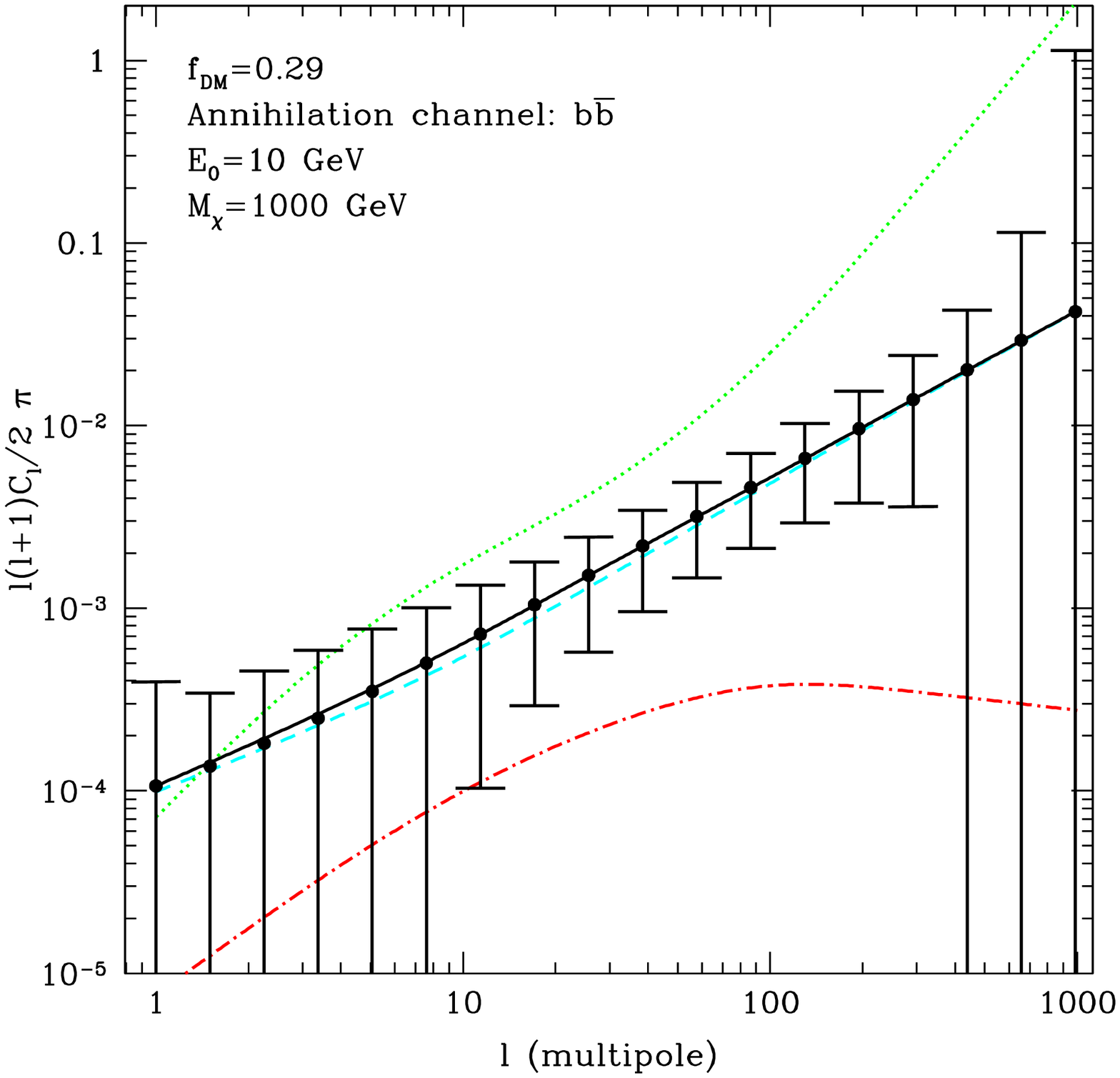}
\caption{Angular power spectrum of the CGB from DM annihilation and
for blazars. Lines are as in Fig.~\ref{fig:clsglast}. Annihilation
channel, energy of detection, DM mass and fractional contribution to
the CGB mean intensity are specified for each panel.}
\label{fig:clsglaststudy}
\end{figure*}

\subsection{Power spectrum dependence on energy of detection, annihilation spectrum and DM mass.}

Even if the results discussed in the section above refer to a
certain specific choice of DM parameters and energy of detection, we
have also repeated the calculations for different cases. Rather than
presenting all the plots, we just show in
Fig.~\ref{fig:clsglaststudy} what we obtained for some benchmarks and
we try to summarize some general guidelines. A more complete analysis,
for example dedicated to the optimization of the energy of detection
as a function of the particle mass, is beyond the scope of this
study.

First we show how our predictions change if we pick another energy
of detection. At energies higher than 10 GeV the galactic foreground
is sensibly suppressed but also the photon number from DM
annihilation is reduced, since the interval of integration in
energy is shrunk. Therefore, it is not trivial to infer which is the
effect on the DM angular power spectrum and its error bars. We find
that even if the CGB mean intensity at $20$ GeV is reduced with
respect to its value at 10 GeV, the power spectrum is increased. We
remind that the power spectrum is normalized to the mean flux, as in
Eq.~(\ref{eqn:Cl}). At an energy of 1 GeV the CGB mean intensity is
comparable with the galactic foreground therefore in
Eq.~(\ref{eqn:Clerror}) we consider $C_N \sim 2 \Omega_{sky}/N_{CGB}.$
For this gamma-ray energy, the signal is sensibly reduced and
the prospects for detection are degraded.

As pointed out in Sec.~\ref{sec:chapter four} for softer energy
spectra, the normalization of angular power spectrum is decreased.
We indeed find this behavior when we compare the results obtained
for the $\tau^+\tau^-$ and for the $b\bar{b}$ DM
annihilation final states. For the ``pessimistic'' case of annihilations into
$\tau^+\tau^-,$ assuming an energy of detection of 10 GeV and
$m_{\chi}=100$ GeV, DM annihilations have to contribute at least
with a fraction $f_{DM}\sim 0.3$ to the mean CGB in order to be
detectable in the CGB angular power spectrum with Fermi.

We finally show the results for  $m_{\chi}=1000$ GeV and $E_0=10$
GeV. The photon spectra from DM annihilation can in good
approximation be scaled with the particle mass defining the
adimensional variable $x=E/m_{\chi},$ as in Fig.~\ref{fig:spectra}.
Therefore, looking at Eq.~(\ref{eqn:Intensity}) and Eq.~(\ref{eqn:Cl}), we
note that there is an approximate scaling which links the
angular power spectra computed at different energies of observations
and for particle of different masses. For example, the choices
($E_0=10$ GeV, $m_{\chi}=1000$ GeV) and ($E_0=1$ GeV, $m_{\chi}=100$ GeV)
actually correspond to the same value of $dN_{\gamma}/dx,$ and should
thus lead to identical  angular power spectra. This scaling is
broken by the dependence of the function
$\tau(z)$ on $E_0$ in Eq.~(\ref{eqn:Intensity}), which
fortunately is not important at the energies of interest, and our
qualitative considerations are still roughly valid. This can be seen
noting that the power spectra in Fig.~\ref{fig:clsglaststudy} for
the two cases above are indeed very similar. Note however that, as already
stressed, different energies of observations significantly affect
the projected errors bars.

\section{Conclusions}
\label{sec:chapter six}

DM annihilation in mini-spikes around IMBHs is a promising scenario for
indirect DM searches with gamma rays. In particular, Fermi is expected to
detect a significant fraction of the IMBH population in the Milky
Way and maybe a few sources in the Andromeda galaxy. The remaining
cosmological mini-spikes will remain unresolved, but could leave
their imprint in the CGB. As shown in \cite{Horiuchi2006}, for a standard neutralino with a mass
$m_{\chi}=100$ GeV and a ``thermal'' annihilation cross section
$(\sigma v) =3$ $10^{-26}$ cm$^3$s$^{-1}$ the predicted CGB flux from
cosmological mini-spikes is comparable to the EGRET CGB flux at gamma-ray energies
of $\mathcal{O} (1-10)$ GeV. We find that, for example, this
corresponds to a fraction $f_{DM}=0.35$ and $f_{DM}=0.72$ of the CGB
at $E=10$ GeV, respectively for DM annihilation into $b \bar{b}$
and $\tau^+\tau^-.$ Fermi is expected to resolve a much larger number of
galactic and extragalactic gamma-ray sources compared to its predecessor EGRET,
with the expectation of reducing the measured unresolved diffuse CGB flux. At the
same time, only IMBHs very close to us will be resolved, therefore
the DM contribution to the  CGB could be increased with respect to
our estimates, based on the mean CGB spectrum extracted from EGRET
data.

However, in absence of characteristic spectral features, it will be
problematic to distinguish DM annihilation and ordinary
astrophysical emissions from the mean GCB intensity. Instead, Ref.~\cite{Ando2006a} showed that gamma-ray anisotropy data could
provide a more suitable tool to pursue this program. In fact,
provided DM annihilation contributes substantially to the CGB mean
intensity, it will be detectable in the CGB angular power spectrum
by Fermi.

Motivated by these considerations, in this paper, we studied the
anisotropies of the CGB in the mini-spikes scenario. Astrophysical
and particle physics uncertainties largely affect the predictions
for the mean CGB intensity from mini-spikes and also the blazar
contribution is currently unknown. Considering these two sources as
the main components of the CGB, we computed their angular power
spectra for different relative contributions and, treating the blazar
component as a known background, we studied the prospects for
DM annihilation detection in the CGB angular power spectrum with
two years of Fermi observations. We expect that considering
unresolved blazars as a background is a reasonable assumption,
since their GLF and bias should be quite reliably reconstructed from the
Fermi source catalog.

We repeated our computations for different detection energies, particle masses and
annihilation modes, showing that our
results are significantly affected by all these parameters.
Interestingly, this could mean that information on these three quantities can actually
be inferred from the measured DM-induced gamma-ray anisotropy power spectrum. However a more detailed
analysis would be necessary to fully study the potential of the
anisotropy technique to reconstruct these parameters. We found
that the shape of the DM power spectrum is very different from that
of blazars, providing a robust handle to disentangle the
two signals. Astrophysical sources other than blazars could
however also contribute to the CGB and, if spatially extended, as clusters of
galaxies, the shape of their angular power spectrum could significantly
differ from that of blazars, which is dominated at large multipoles
by the Poisson term. We stress that even in this case we could
calibrate the astrophysical power spectrum at low energies, where DM
annihilations are negligible and subtract it from the measured total
CGB power spectrum at the energies of interest. In fact, for sources
with power-law energy spectra, the gamma-ray angular power spectrum
is energy independent and this condition is common to almost any class of standard
astrophysical gamma-ray emitter.

In conclusion, we showed that the prospects for detecting DM annihilation
from cosmological mini-spikes in the angular CGB power
spectrum with Fermi are promising, and that the analysis of the
anisotropy power spectrum allows not only a discrimination of a DM component
against astrophysical sources, but also a better understanding
of the structures where the DM signal originates.

\acknowledgments
MT thanks the California Institute of Technology
for hospitality and partial support during the preparation of this
work. SA is supported by the Sherman Fairchild Foundation. SP is
supported by the US Department of Energy under grant
DE-FG02-04ER41268 and by the National Science Foundation.

\addcontentsline{toc}{chapter}{Bibliografia}
\bibliographystyle{apsrev}
\bibliography{draft4}



\end{document}